\documentclass[structabstract,twocolumn]{aa}
\usepackage{graphicx}
\newcommand{\vsini}[1]{$v\cdot\sin(i)$}
\newcommand{\teff}[1]{$T_{\rm eff}$}
\def\cm1{$\rm cm^{-1}$}

\def\DE{D\kern-0.75em \raisebox{1.0pt}{=}\ }
\def\GR{\cal \char'122}

\def\Sum{N_{\rm tot}}
\def\(({\left (}
\def\)){\right )}

\begin{document}
%

%
   \title{Abundances in the Herbig Ae star HD 101412\thanks{
Based on observations
obtained at the European Southern
Observatory, Paranal and La Silla, Chile
(ESO programmes 077.C-0521(A), 081.C-0410(A) and
383.C-0684(A)}}

   \subtitle{Abundance anomalies; $\lambda$ Boo--Vega characteristics?}

   \author{C. R. Cowley
          \inst{1}
          \and
          S. Hubrig\inst{2}
          \and
          J. F. Gonz\'{a}lez\inst{3}
          \and
          I. Savanov\inst{4}
          }

   \offprints{C. R. Cowley}

   \institute{Department of Astronomy, University of Michigan,
              Ann Arbor, MI 48109-1090, USA\\
              \email{cowley@umich.edu}
         \and
             AIP, An der Sternwarte 16, 14482 Potsdam, Germany\\
              \email{shubrig@aip.de}
          \and
             Instituto de Ciencias Astron\'{o}micas, del la Terra y del Espacio,
             Casilla 467, 5400 San Juan, Argentina  \\
             \email{fgonzalez@icate-conicet.gob.ar}
          \and
             Institute of Astronomy, Russian Academy of Sciences,
             Pyatnitskaya 48, Moscow 119017, Russia
             \email{isavenov@rambler.ru}}   

   \date{Received month date, 2010/  Accepted month date year}

\abstract
{Recent attention has been directed to
abundance variations among very young stars.}
{To perform a detailed abundance study 
of the Herbig Ae star HD 101412, taking advantage of its
unusually sharp spectral lines.}
{High-resolution spectra are measured for accurate
wavelengths and equivalent widths.  Balmer-line fits and
ionization equlibria give a relation between $T_{\rm eff}$, and
$\log(g)$.  Abundance anomalies and uncertain reddening
preclude the use of spectral type or photometry to fix
$T_{\rm eff}$.  
Excitation temperatures are used to break the degeneracy
between $T_{\rm eff}$ and $\log(g)$.}
{Strong lines are subject to an anomalous saturation
that cannot be removed by assuming a low microturbulence.  By
restricting the analysis to weak ($\le 20$m\AA) lines, we 
find consistent results for neutral and ionized species,
based on a model with $T_{\rm eff} = 8300$K, and $\log(g)=3.8$.
The photosphere is depleted in
the most refractory elements, while volatiles are normal
or, in the case of nitrogen, overabundant with respect
to the sun.  The anomalies are unlike those
of Ap or Am stars.}
{We suggest the
anomalous saturation of strong lines arises from heating
of the upper atmospheric layers by infalling material
from a disk.
The overall abundance pattern  may be related to those
found for the $\lambda$ Boo stars, though the depletions
of the refractory elements are milder, more like those of Vega.
However, the intermediate volatile zinc is depleted, precluding
a straightforward interpretation of the abundance pattern
in terms of gas-grain separation.}  
\keywords{Stars: chemically peculiar -- Stars: abundances
 -- Stars: individual (HD101412)-- Stars:pre-main-sequence}
\maketitle 
  \titlerunning{HD 101412: refractory depletions}
  \authorrunning{Cowley, Hubrig, Gonz\'{a}lez, Savanov}

\section{Introduction}
The star HD 101412 (CD $-$59 3865, V1052 Cen) belongs to the
group of Herbig Ae/Be stars, whose  members are considered
more massive counterparts of T Tauri pre-main sequence stars.
Masses range from 2-10 $M_\odot$ (Hern\'{a}ndez, et al.
2005).
Their spectral energy distribution is characterized by the
presence of an infrared excess due to thermal
re-emission of circumstellar dust, which is thought
to be the signature of a circumstellar disk
(cf. Hartmann 2009, Sect. 8.8).
According to Fedele et al. (2008) HD 101412 is a
group II source (using the classification of Meeus
et al. (2001), with the flat disk self-shadowed by dust.
Wade, et al. (2005) discussed HD 101412 as a possible
progenitor of magnetic chemically peculiar (CP2 or Ap)
stars.  A number of recent studies have
been devoted to this star (cf. Hubrig, et al. 2009, 2010),
including an estimate of its metallicity (Guimar\~{a}es,
et al 2006).  The absorption lines are unusually sharp,
leading to suggestions that the star may be seen nearly
``pole on.'' Intrinsically slow rotation cannot
be excluded.  Indeed, Fedele, et al. (2008) propose a
model with a disk inclined by 20$^\circ$ to the
line of sight, a model now supported by a newly-found
photometric period.

Very recent work shows that HD 101412 is a low-amplitude
photometric variable
with the period of $42\fd 0$
(Mikul\'a\v sek et al., \textit {in preparation}).
If the period is a rotational one, it then comports well
with our spectroscopic value, $v\cdot\sin i=3\pm 1$
km s$^{-1}$, and excludes a
conclusion that the narrow
lines could {\it only} be due to a (nearly) pole-on viewing angle.


Numerous emission features in Herbig Ae stars indicate
an atmospheric structure that must differ from the
classical, one-dimensional models typically used for
abundance analyses.
Nevertheless, abundance work on
these objects has been attempted (cf. Guimar\~{a}es,
et al. 2006, Acke and Welkens 2004).  The latter
study is of particular interest to the results presented
here, as the authors sought the signature of the
$\lambda$ Boo phenomena, citing the seminal paper by
Venn and Lambert (1990).  These authors noted that the
most volatile elements, C, N, and O showed mild 
abundance depletions,
while the more refractory elements showed significant
depletions, 1 dex or more.  They noted the similarity
of these depletions to those of the interstellar gas
which is well correlated with condensation temperatures
(cf. Sect.~\ref{sec:abundances}).

The sharp spectral lines of HD 101412 make it ideal
for abundance work.  Although a magnetic field has
been detected (cf. Wade, et al. 2005),
magnetic enhancement of the lines is
masked by other factors, to be discussed
below (Sect.~\ref{sec:anomsat}).
Hubrig et al. (2010) noted that the Zeeman patterns
of several lines were resolved, or partially resolved.
Nevertheless, most atomic lines are sufficiently
sharp that numerous relatively unblended lines could
be found for abundance work.






\section{Spectra}
\label{sec:spec}

The current paper is based on a subset of the spectra
described by Hubrig et al. (2010).  Most equivalent
widths were measured on
two HARPS spectra (averaged) obtained in 2009 on
23:47 UT  of 4 July, and 00:13 UT of 5 July.
About 10 per cent of the equivalent widths are
from UVES spectra obtained on 9 (visual and IR) and
15 April (UV) 2009.

Wavelengths were measured on the HARPS spectrum in the range
$\lambda\lambda$3782-6911. 
The averaged spectra
were subjected to mild Fourier filtering.
Signal-to-noise estimates, made directly from
the filtered spectra average 148.  HARPS spectra
have a resolving power (RP) of 120000, but
averaging and filtering reduce this to an estimated
40000.  Fortunately, equivalent widths are independent
of RP;  we exchange RP for noise reduction.

We have used accurate wavelength measurements both
for line identification, and to find relatively
unblended lines.

The UVES spectrum obtained on 15 April 2009,
was measured for wavelengths in the range
$\lambda\lambda$3301--4517.  A few equivalent
widths below the Balmer
Limit were obtained from this spectrum.

Additional wavelengths were measured from UVES
spectra taken on 9 April 1999
($\lambda\lambda$6911--7519 and 7661--9461).
The Ca {\sc ii} infrared triplet was 
examined for possible isotopic shifts discovered
by Castelli and Hubrig (2004).  The lines were shifted
by 0.02 to 0.03~\AA\, with respect to the
solar wavelengths of these lines, a rather
small shift that could be entirely due to measurement
error.

Hubrig, et. al. (2010) demonstrate equivalent width
changes of the order of 20\% (0.08 dex).  The
variations were correlated with an assumed phase,
13\fd 86 days nearly one third the newly determined period
of 42\fd 0.  Also possible, are line strength
changes due to time-dependent motions of the circumstellar
material.  These variations must
be kept in mind in evaluating the abundances which were
determined here primarily from a single phase.  An
error of 0.08 dex is marginally significant, relative
to other uncertainties in the abundances
(cf. Table~\ref{tab:abtab}).

Balmer line profiles were measured on a FORS\,1 spectrum
obtained on 22 May 2008, in visitor mode at ESO.
We used the (200kHz, low, 1$\times$1) readout mode, which
makes it possible to achieve a S/N  ratio of about
1000--1200 with a single exposure and
the GRISM\,600B in the wavelength range 3250--6215\,\AA{}
to cover all hydrogen Balmer lines from H$\beta$ to the
Balmer jump.  A slit width of 0$\farcs$4 was used to
obtain a spectral resolving power of $R\approx2000$.
Details of data reduction are given by Hubrig et al. (2004).

\section{Model atmospheres and spectral calculations\label{sec:modcal}}

All model atmospheres are based on $T-\tau$ relations
from the version of ATLAS9 (Kurucz 1993) of
the Trieste group (cf. Sbordone, et al. 2004).
Given
$T(\tau)$, the depth integrations from
$\log(\tau_{5000})$ of $-5.4$ to $+1.4$ were carried
out with software and opacity routines written and
used at Michigan for several decades.  Agreement with
ATLAS9 Models posted on Castelli's (2010) site
are excellent (see Castelli \& Kurucz 2003).   Spot
checks using WIDTH9 for weak Fe {\sc i} and {\sc ii} lines,
which are independent of the different damping
constants of the two codes, show
agreement within 0.02 dex.  Weak lines have
equivalent widths of 20 m\AA\, or less.

\section{Fixing the atmospheric parameters\label{sec:fixatm}}
\subsection{Colors and spectral type}
Broad and intermediate-band photometric measurements
are typically used to fix the effective temperatures prior
to an abundance analysis.
Johnson (Vieira, et al. 2003) and Geneva (Mermillod,
Hauck, and Mermillod 2007) indices are available
for HD 101412; we found no Str\"{o}mgren photometry. 
Unfortunately, the reddening correction
is most uncertain.  For example, Vieira, et al.'s $U-B=0.15$,
$B-V=0.18$ do not lead to an
intersection with a standard $U-B$ vs.
$B-V$ plot for either main sequence stars (Cox 1995,
Table 15.7) or a ZAMS (Table 15.9).
Since the star is known to be peculiar
and variable, currently available photometry
is not adequate
to fix the effective temperature.

The spectral type of HD 101412 is
given variously as B9-A0 V-III (e.g.
B9/A0 V, Houk and Cowley 1975).
According to Cox (1995), luminosity
class V stars have effective temperatures
in the range 10500 to 9790K, while we conclude
$T_{\rm eff} \approx 8300$K.
The spectral type of normal stars of this class
depend strongly
on the Ca {\sc ii}
K-line, or rather its strength relative to Ca {\sc ii} H
+H$\epsilon$.
The intensity of the Ca {\sc ii} lines would
be significantly diminished by the $\approx 0.6$
dex underabundance we find for this element.

We
conclude that the spectral type is not a useful
indicator of the temperature of the star, or
of the total absorption, $A_V$.

\subsection{The Balmer lines\label{sec:balmer}}

\begin{figure}
\resizebox{\hsize}{!}{\includegraphics[angle=-00]{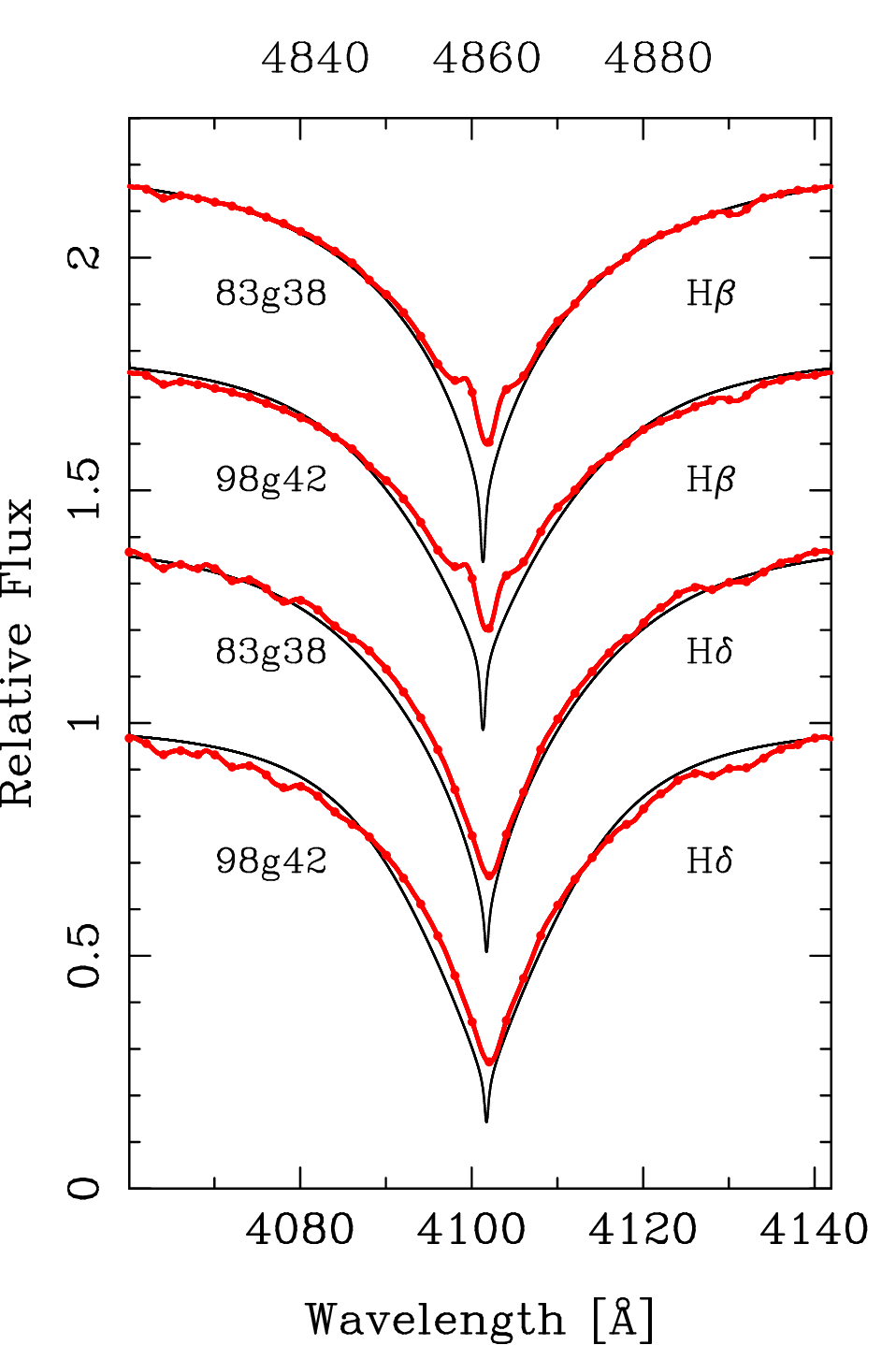}}
\caption{ H$\beta$ (above, top wavelength scale) and H$\delta$ profiles
(bottom wavelength scale) for $T_{\rm eff} = 8300$K, $\log(g)=3.8$ and 
9800K, $\log(g) = 3.8$.  Observed profiles are in gray (red in online
version) with dots.  The fits for the lower temperature are as good
or arguably better than at the higher temperature. The relative flux
scale is correct for the lowest profile pair.  Upper pairs are successively
raised by 0.4 units.
\label{fig:balmers}}
\end{figure}

Balmer line profiles are difficult to work with
on high-dispersion material.  This is especially true
for echelle spectra, as the broad lines typically
span echelle orders.
We have attempted to fit profiles
from an averaged HARPS spectrum, but have relied
primarily on the FORS1 spectrum (Sect.~\ref{sec:spec})
which included H$\beta$ and higher members
of the Balmer series (Fig.~\ref{fig:balmers}).

Guimar\~{a}es, et al. (2006) made use of Balmer lines
but did not discuss $T_{\rm eff}-\log(g)$
degeneracy.  They
conclude $T_{\rm eff} = 10000\pm 1000$K
(see their Table 2).
In Hubrig, et al. (2009),
only the wings of H$\beta$ were used to find the
atmospheric parameters, $T_{\rm eff} = 10 000$,
$\log{(g)} = 4.2-4.3$.  The authors note that a 9000K,
$\log{(g)} = 4.0$ model also gave a good fit to the
H$\beta$ profile, but that with these parameters
``many narrow lines'' were seen in the synthetic
spectrum that did not appear
in the observed spectrum.  These lines might not
have appeared in the calculated spectrum if the
abundances were low, as we currently assume.

The Hubrig et al. (2009) results show that within the
relevant temperature range for HD 101412, the
Balmer profiles are sensitive to both temperature
and gravity.  Our current calculations show
that quite good fits to the Balmer profiles may
be obtained for a temperature as low as 8300K,
if a $\log{(g)} = 3.8$ is assumed
(see Fig.~\ref{fig:balmers}).
A still lower temperature could be accommodated
by assuming an even lower gravity.

Wavelength measurements of the cores of the Balmer lines on
the FORS1 spectra show systematic red shifts that
increase systematically from about +9 at H19 to
+23 km s$^{-1}$ at H$\beta$.
Core shifts are marginally
detectable in Fig.~\ref{fig:balmers}

\subsection{Ionization equilibrium\label{sec:feeqlib}}

We assume here that an optimum model will yield the
same abundances from two stages of ionization of
a given element.
In a preliminary approach to obtain model
parameters, we
adopted a grid of temperatures and gravities shown
in Table\,~\ref{tab:grid}.  The grid was chosen
to bracket the expected stellar parameters.
Provisional abundances
were then calculated from equivalent width measurements
of Fe {\sc i} and {\sc ii} lines using models with the parameters
given in the table.

\begin{table}
\caption{Provisional abundances ($\log(Fe/\Sum)$)
from Fe {\sc i} and
Fe {\sc ii} for various temperatures and surface gravities.
Nearly equal abundances for Fe {\sc i} and {\sc ii}
are in bold face.\label{tab:grid}}
\begin{tabular}{ccccccl}   \hline
$\log{(g)}$&\multicolumn{5}{c}{$T_{\rm eff}$(K)}&Spec. \\ \hline
      &7800   &8300   &8800   &9300  &9800    &  \\  \hline
4.2   &$-5.37$&$-5.14$&{\bf $-$4.78}&$-4.40$&$-4.04$&Fe {\sc i} \\
4.2   &$-5.00$&$-4.97$&{\bf $-$4.84}&$-4.76$&$-4.57$&Fe {\sc ii} \\ \hline
3.8   &$-5.38$&{\bf $-$5.09}&$-4.67$&$-4.28$&$-3.92$&Fe {\sc i} \\
3.8   &$-5.17$&{\bf $-$5.10}&$-4.91$&$-4.83$&$-4.73$&Fe {\sc ii} \\ \hline
3.2   &{\bf $-$5.37}&$-4.96$&$-4.50$&$-4.09$&$-3.72$&Fe {\sc i} \\
3.2   &{\bf $-$5.32}&$-5.18$&$-5.00$&$-4.95$&$-4.83$&Fe {\sc ii} \\ \hline
\end{tabular}
\end{table}

\begin{figure}
\resizebox{\hsize}{!}{\includegraphics[angle=-90]{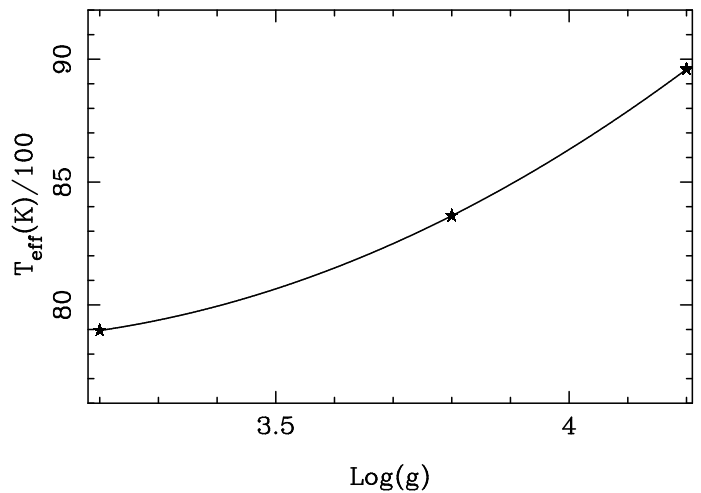}}
 \caption{Values of $T_{\rm eff}/100$ and $\log{(g)}$
 yielding equal abundances for Fe {\sc i} and Fe {\sc ii}.
 \label{fig:pltg}}
 \end{figure}

A relation between \teff\, and $\log{(g)}$ is
obtained as follows.
For each of the three values of $\log{(g)}$, we plot
$\log(Fe/\Sum)$ vs. \teff, and find a value of \teff\,
where the abundances from Fe {\sc i} and Fe {\sc ii}
agree.  This is illustrated in Fig.~\ref{fig:pltg}.

\section{Anomalous saturation}
\label{sec:anomsat}
The abundance from a set of equivalent widths of a given
spectrum, e.g. Fe {\sc i}, or Ti {\sc ii}, should not,
in principle, depend on the line strength.  However,
it has been known since the early days of curve of growth
analysis, that drifts of abundance with line strength
are common.  They can result from a variety of causes
ranging from errors in the oscillator strengths to
equivalent width measurements.  In a common situation,
strong lines will give a higher abundance than weak ones,
and the abundance worker can remove the trend by assuming
an additional source of broadening known as
microturbulence ($\xi_t$).  This broadening arises from gas motions
due to convection, mixing, or other sources not included
in the basic expression for the absorption profile,
such as hyperfine structure or Zeeman effect.

\begin{figure}
\resizebox{\hsize}{!}{\includegraphics[angle=-90]{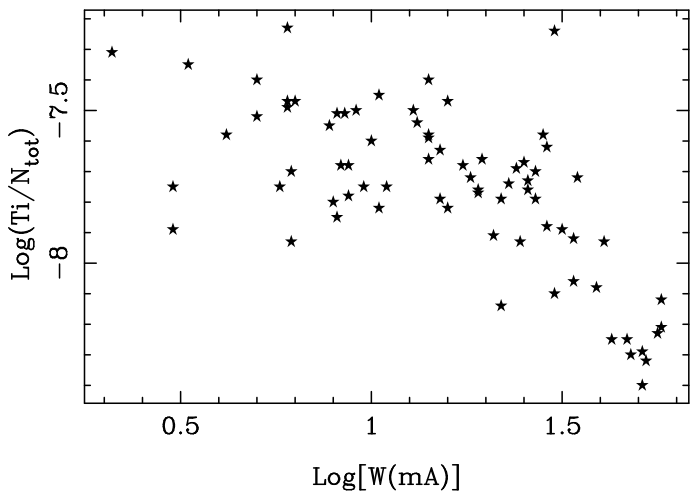}}
\caption{$\log{(Ti/\Sum)}$ vs. $\log{[W_\lambda(mA)]}$ for
 Ti {\sc ii} lines using the model with
$T_{\rm eff} = 8300$K, $\log(g)=3.8$.  A microturbulence
$\xi_t = 0$, was assumed.  The downward trend
of abundance with equivalent width illustrates the
anomalous saturation.  (Additional plots may be
found at the URL given in Sect.~\ref{sec:abundances}.
\label{fig:ti2aw}}
\end{figure}

Plots of abundance vs. equivalent width for HD 101412
slope downward {\it even when no microturbulence is assumed.}
Fig.~\ref{fig:ti2aw} shows the behavior for Ti {\sc ii}
lines.  Not all plots show the flattening for weak lines.
In particular, the Fe {\sc ii} plot does not. However, most do.

An obvious mechanism that would weaken lines in young
stars is veiling.  This arises from excess continuum
radiation arising from some aspect star-formation
process (cf. Hartmann 2009). It is important to assess
the possible effects of veiling, which would weaken
the absorption lines and mimic lower abundances.


We believe the effect of veiling to be
minor, for the following reasons:

\begin{itemize}
\item[$\bullet$] Carbon, oxygen, and sulfur give abundances
close to solar.  In the case of nitrogen, we find a
significant excess.  Veiling, if it were present, should
have affected lines from these elements.
\item[$\bullet$] Veiling is often wavelength dependent.  There
is no indication of a drift in our abundances with wavelength
for the five spectra having lines on either side of the Balmer jump:
Mg I, V II, Cr I, Co I, and Co II.
\item[$\bullet$] Veiling is caused by an excess continuum due to
infall from a disk.  The veiling is less
important for the Herbig Ae stars, relative to the
T Tauri or weak T Tauri stars.  There is no indication
of X-rays from HD 101412, which would accompany
significant infall of material.
\end{itemize}

\begin{figure}
\resizebox{\hsize}{!}{\includegraphics[angle=-90]{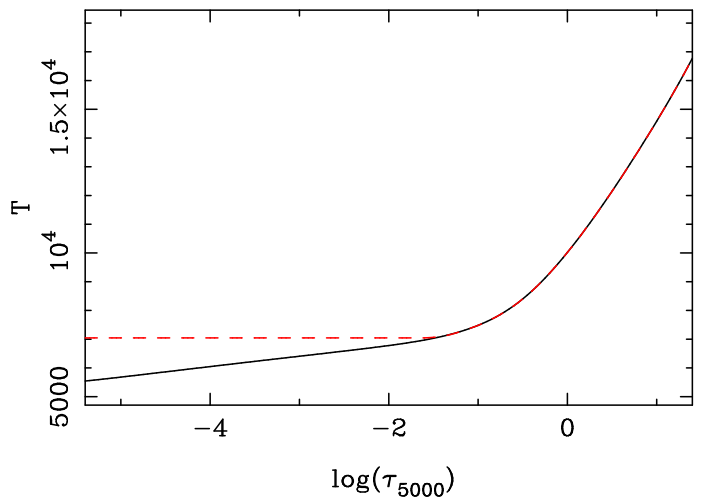}}
\caption{Temperature distributions for the ATLAS9 model
with $T_{\rm eff} = 8300$K, $\log(g)=3.8$ (solid line), 
and a ``hot top'' model (dashed) which gives no anomalous 
saturation for $\xi_t = 0$.}  
\label{fig:hottop}
\end{figure}

We have found that anomalous saturation
does not occur in a model with a hot upper atmosphere.
If we arbitrarily raise the temperature
from the standard ATLAS9 $T(\tau)$ the effect vanishes
(see Fig.~\ref{fig:hottop}).
Temperature changes were made by trial and error until the
anomalous saturation was no longer evident in a plot
of abundance from Fe II lines vs. equivalent width.
The altered model is surely not unique.
Since the results are from an LTE calculations, we take
it as indicative of the problem with the ATLAS9 model for
HD 101412 specifically.
A realistic pursuit of this problem is best left to a
NLTE study.

\section{ $T_{\rm eff}-\log{(g)}$ degeneracy; 
curves of growth\label{sec:tgdeg}}

We can see from Fig.~\ref{fig:pltg}, and from
Table~\ref{tab:grid}, that higher temperatures require
higher gravities.  This is true for both the iron equilibrium and the
Balmer profiles.  The
temperature (9500-10000K) obtained by earlier workers
would not yield acceptable Fe ionization,
though the Balmer lines could be accommodated with a
higher surface gravity.

This leaves the question of how to break the degeneracy
between temperature and gravity.  In one of the relatively
small number of abundance analyses of pre-main sequence
stars, Acke and Waelkens (2004) determined atomic
excitation temperatures ``for ions with many observed
lines."
Specifically, they required that the effective temperature
of the model be such that there is no drift in the abundance
determined from individual lines with excitation potential.

We have attempted to use the Acke and Waelkens (2004) method. 
Our results marginally favor a 8300K--$\log(g)=3.8$ model
over a 9800K--$\log(g) = 4.2$ model, though not decisively.
The wrong temperature should show a clear difference between
abundances from low- and high-excitation lines. Unfortunately
anomalous saturation precluded use of many stronger lines,
which are typically of low excitation.  As we shall see, this
was not a problem with the second method, based on a curve
of growth technique.

Since our depth-dependent models are not accurate for
this Herbig Ae star, it is
worthwhile to explore results of a more basic approach
where the photosphere is approximated by a uniform slab,
with a single,
mean value of the temperature and pressure. 
This simple technique was once widely used 
in the chemical analysis of stellar spectra, employing what were called 
Schuster-Schwarzschild (henceforth, SS) models. (cf. Aller 1963).
The basic method is still routinely employed when the
conditions along the line of sight are not well determined
(cf. Spitzer 1978,
Rachford, et al. 2001, Hobbs 2005).  
This is the case for the upper layers of HD 101412.  

Note: we use the SS models 
{\it only} to find an excitation temperature and break the
$T_{\rm eff}-\log(g)$ degeneracy.

To determine an excitation temperature, we plot
$\log{(W_{\lambda}/\lambda)} + 6$ vs.
$\log{(gf\lambda)} - \theta\cdot\chi$.  Here, 
$\theta = 5040./T_{\rm exc}$, and $\chi$ is the lower excitation
potential in eV.
One adjusts the value of $\theta$ until the points belonging
to lines with different excitation delineate, as well as possible,
the same curve.

Constants are added to
both the ordinates and abscissae of the {\it observed}
points, in order to make
the theoretical and stellar curves of growth overlap.
The ordinate of
the theoretical curve is 
$\log[W_\lambda/(2r_0\Delta\lambda_D)]$, while that
of the empirical plot is $\log(W_\lambda/\lambda)+6$.
The vertical shift necessary to superimpose the two
curves contains information on the
Doppler width,
$\Delta\lambda_D = \frac{\lambda}{c}\sqrt{2{\GR}T/\mu + \xi_t^2}$,
and the maximum line depth $r_0$,
discussed in more detail below.
The horizontal shift contains information on the column density,
but that is not needed here.
The analytical curves were
originally due to van der Held (1931), and are
are shown in the figure
as solid lines
for several values of the ratio
of $a = \gamma_\lambda/(2\cdot \Delta\lambda_D)$.
Here $\gamma_\lambda$ is the damping constant
(in cm),
$\mu$ is the molecular (atomic) weight, and
${\GR}$ the gas constant.
Note that
a constant has been added to our abscissa so that
$X$ and the ordinates
of the theoretical curves are the same for {\it weak}
lines.
We made curves of growth for Ti {\sc ii}, 
Fe {\sc i}, and Fe {\sc ii},
spectra with better-quality oscillator strengths, 
and Ca {\sc i}
We have tried to select oscillator strengths of optimum
quality. Sources are:

\begin{itemize}
\item[$\bullet$] Ti {\sc ii}.  Oscillator strengths are from Pickering,
Thorne \& Perez (2001)
\item[$\bullet$] Cr {\sc i}.  Oscillator strengths from Sobeck, Lawler,
\& Sneden (2007).
\item[$\bullet$] Cr {\sc ii} Oscillator strengths from Nilsson, et al.
(2006) or VALD (Kupka, et al. 1999)/Kurucz (1994) taking
only LS-permitted lines.
Results were essentially the same.
\item[$\bullet$] Fe {\sc i}: Lines from the NIST site (Ralchenko, et
al. 2010) with accuracy B+ (7 lines) and C+ (11 lines).
Unfortunately, there are no lines with excitation potentials
above 2.6 eV of accuracy B+ (see the NIST site for an
explanation of ``accuracy.'')
\item[$\bullet$] Fe {\sc ii}: We adopt the oscillator strengths of
Mel\'{e}ndez and Barbuy (2009).
\end{itemize}

\begin{figure}
\resizebox{\hsize}{!}{\includegraphics[angle=-90]{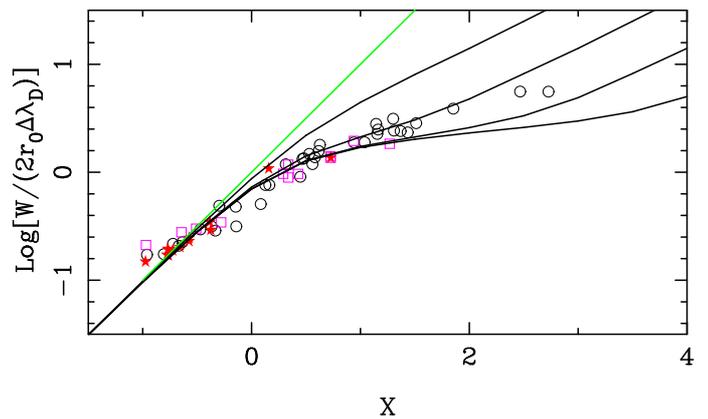}}
\caption{Curve of growth for Fe {\sc ii} lines using
oscillator strengths from Mel\'endez and Barbuy (2009).
Unit slope is indicated by the solid gray line.
Along this line, the ordinate, and the abscissa ($X$),
are equal.  SS model
curves are drawn for
$a = \gamma_\lambda/(2\cdot \Delta\lambda_D)$ of 1, 0.1, 0.01,
and 0.001 (from the top).  The abscissae for the points
are
$\log{(gf\lambda)} - \theta\cdot\chi$.
Open circles are for $2.58\le\chi\le 2.89$, open squares
for $3.20\le\chi\le 3.89$, and filled stars for
$5.51\le\chi\le 6.22$ eV.  The observed points have been made
to coincide by adding +5.7 to the abscissa, and $-$0.6 to
the ordinate.  The latter figure is relevant for the
microturbulence.  This plot was made with $\theta = 0.7$
(7200K).\label{fig:fe2aap}}
\end{figure}

Fig.~\ref{fig:fe2aap} shows a curve of growth
for Fe {\sc ii}.  Too large a $\theta$
(too low a temperature) will move the filled
stars (highest $\chi$) to the left with respect to the
open circles (lowest $\chi$).  Too small a $\theta$,
would move the highest-$\chi$ points to the right.
Points for the lowest-excitation lines move relatively
little.

\begin{figure}
\resizebox{\hsize}{!}{\includegraphics[angle=-90]{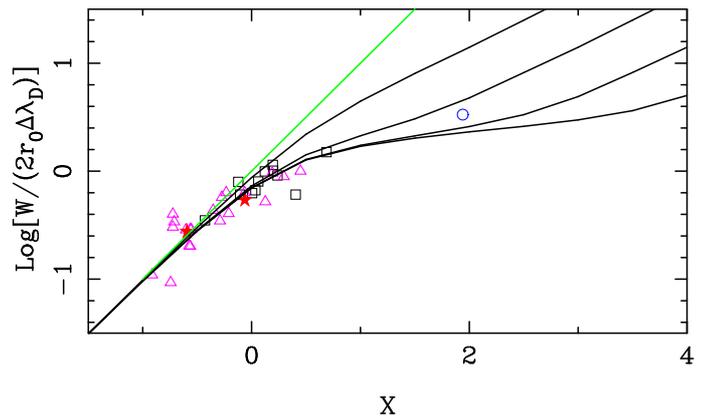}}
\caption{Curve of growth for 35 Ca {\sc i} lines.
The resonance line,
$\lambda$4227 is the open circle.  Open squares are
for the range $1.88\le\chi\le 1.90$, 
triangles $2.50\le\chi\le 2.71$,
and filled stars (2 lines) 2.93 eV.  $\Delta x = \,$ 1.7
$\Delta y = -0.65$ and $\theta =\, $ 0.68 ($T = 7410$K).
Theoretical curves as in Fig.~\ref{fig:fe2aap}}
\label{fig:ca16565}
\end{figure}

In the case of Ca {\sc i}, there is no good recent set of
oscillator strengths, apart from the three lines
given by Aldenius, Lundberg, and Blackwell-Whitehead
(2009), $\lambda\lambda$6162.2, 6122.2, and 6102.7).
These three lines are all of moderate strength
(9.6 to 25.3 m\AA), and give $-$6.41, $-$6.42, and $-$6.30
for $\log(Ca/\Sum)$.  We adopt $-$6.38; the mean
of values from 35 lines is $-6.43\pm 0.23$\,sd.
A curve of growth is shown for Ca {\sc i} in
Fig.~\ref{fig:ca16565}. Oscillator strengths for
the remaining 32 lines are from the NIST site
(Ralchenko, et al. 2010).

Additional curves of growth and examples showing results
of a wrong temperature may be found at the url given
in Sect.~\ref{sec:abundances}.

Curves of growth provide a measure of the excitation
temperature, the microturbulence ($\xi_t$),
and the damping constant.
The excitation temperatures found from the curves
of growth,
corresponding to $\theta = 0.65$ to 0.7 ($T = 7750$ to
7200K) are much cooler than any temperature discussed so far
for HD 101412.  However, the excitation temperature
from the line spectrum is
expected to be significantly lower than that
from the continuum.  We have made calculations based
on model atmospheres to verify that the above excitation
temperatures are to be expected for an
atmosphere with $T_{\rm eff}=8300$K.

The vertical displacement ($\Delta y$) necessary to superimpose the
theoretical and empirical curves gives a relation between
the maximum line depth $r_0$, an assumed
relevant temperature, and the microturbulence. 

Our problem with the SS-model curves of growth is essentially
the same as that with the ATLAS9 model.  The maximum line 
depth expected for either model is too large, relative to the value 
needed in HD 101412 to
give a finite microturbulence.  In the case of the numerical model,
$r_0$ in LTE is simply
$[1-B_\lambda(\tau_{min})/(2F_0^c)]$.  The value,
at $\lambda$4481 in the 8300K$-\log(g)=3.8$ model is 0.90.  
For the SS model,
if we adopt $r_0 = 0.53$ from the observed $\lambda$4481 cores,
and an excitation temperature of 7200K, we find
 $\xi_t^2 < 0$.


The strong lines in Fig.~\ref{fig:fe2aap} fall between
theoretical curves for the parameter
$a = \gamma_\lambda/(2\cdot \Delta\lambda_D$) of 0.01 and 0.1.
Note that $\gamma$ (without a subscript) traditionally
means $\gamma_\omega$, or
the FWHM of a dispersion profile expressed as a function
of $\omega = 2\pi c/\lambda$.

Using $T = 8300$K, and a mean $\lambda = 4.6\cdot 10^{-5}$cm,
we obtain $\gamma = 4.29\cdot 10^9$ s$^{-1}$ for $a = 0.1$.
This is 41 times the classical damping constant
$\gamma_{\rm cl} = 0.22/\lambda^2$ s$^{-1}$ (with $\lambda$ in cm).
For
$a = 0.01$, the result is 4 times the classical.  A
value commonly cited for the empirical
damping constant is 10 times the
classical value (cf. Mihalas 1970).
Cowley and Cowley (1964)
obtained $\gamma/\gamma_{\rm cl} = 25$ for the sun.
We thus see the value from this crude curve
of growth is entirely reasonable for a star on 
or near, the main sequence.

\section{Abundances}
\label{sec:abundances}
All abundances were determined from depth-dependent
model calculations of equivalent widths
assuming zero microturbulence.
Oscillator strength references not explicitly
mentioned
are from NIST (Ralchenko, 2010, preferred) or
VALD (Kupka et al. 1999), except for
Zr {\sc ii}, where the values are from Ljung, et al.
(2006).

Plots of abundance vs. $\log(W_\lambda)$ were
made for all species with more than three lines.
Outliers could nearly always be understood and
either corrected or dropped because of errors,
such as misidentifications or blends.  
Whenever plots of abundance vs. equivalent widths
exhibited a downward trend for stronger lines,
abundances were determined from lines with
equivalent widths $\le 20$ m\AA.

Additional plots were routinely 
made of abundance vs. excitation potential, and
equivalent width vs. excitation potential.
Our restriction to weak lines only for abundances
eliminated most systematic effects, though
some small effects are inevitable.
Additional plots, and detailed line-by-line
results are available from the first author
or at the url:{\bf
\newline  www.astro.umich.edu/{$\sim$}cowley/hd101412/}

Abundances are summarized in Table~\ref{tab:abtab}.  Errors
in Col 3 are standard deviations for the number of lines used,
or the difference in abundances when only 2 lines were available.
For V {\sc i}, the error is the difference from a single V {\sc i}
line, and the mean for V {\sc ii}.

No exotic elements, were identified.  Barium is the heaviest
element positively identified, and the abundances listed for
the lanthanides cerium and europium are upper limits.  There
was no indication of gallium or the heavier noble gases.

\begin{table}
\caption{Adopted abundances.  Upper limits assume
a 1 m\AA\, line for Ce {\sc ii} $\lambda$4686,
and Eu {\sc ii} $\lambda$4205.
The 50\% Condensation temperatures (50\%$T_C$) in the table
are from Lodders (2003).
Elements with low values of
50\%$T_C$ are volatile.  Solar abundances are from
Asplund, et al. (2009).\label{tab:abtab}}
\begin{tabular}{l c c r c c} \hline
Spec  &$\log(El/\Sum)$ & sd.  & No.   & Sun&50\%$T_C$(K)\\ \hline
C {\sc i} & $-$3.70  & 0.26  & 10  & $-$3.61& 40\\
N {\sc i} & $-$3.49  & 0.26  & 10  & $-$4.21& 123\\
O {\sc i} & $-$3.21  & 0.19  & 11  & $-$3.35& 179\\
Na {\sc i} & $-$5.94  & 0.09  &  4  & $-$5.80& 953\\
Mg {\sc i} & $-$5.00  & 0.24  & 11  & $-$4.44& 1327\\
Mg {\sc ii} & $-$5.06  & 0.19  &  8  &        &    \\
Al {\sc i} & $-$6.22  & 0.56  &  5  & $-$5.59& 1641\\
Si {\sc i} & $-$4.92  & 0.40  &  9  & $-$4.53&  1302\\
Si {\sc ii} & $-$5.30  & 0.31  & 10  &        & \\
S {\sc i} & $-$4.96  & 0.10  &  8  & $-$4.92&655 \\
Ca {\sc i} & $-$6.38  & 0.07  &  3  & $-$5.70& 1505\\
Ca {\sc ii} & $-$6.30  & 0.52  & 10  &        &  \\
Sc {\sc ii} & $-$9.36  & 0.09  & 11  & $-$8.98& 1647\\
Ti {\sc ii} & $-$7.62  & 0.16  & 43  & $-$7.09& 1573\\
V {\sc i}  & $-$8.29  & 0.19  &  1  & $-$8.11& 1427 \\
V {\sc ii}  & $-$8.48  & 0.22  & 28  &        &  \\
Cr {\sc i} & $-$6.89  & 0.26  & 17  & $-$6.40& 1427\\
Cr {\sc ii} & $-$6.95  & 0.16  &  6  &        & \\
Mn {\sc i} & $-$6.98  & 0.32  & 12  & $-$6.61& 1150\\
Mn {\sc ii} & $-$6.95  & 0.13  &  7  &        & \\
Fe {\sc i} & $-$5.07  & 0.19  & 18  & $-$4.44& 1328   \\
Fe {\sc ii} & $-$5.03  & 0.17  & 30  &        & \\
Co {\sc i} & $-$7.62  & 0.17  & 17  & $-$7.05& 1347 \\
Co {\sc ii} & $-$7.46  & 0.24  &  6  &        &    \\
Ni {\sc i} & $-$6.34  & 0.21  & 40  & $-$5.82& 1348 \\
Ni {\sc ii} & $-$6.12  & 0.27  &  2  &        &   \\
Zn {\sc i} & $-$8.54  & 0.12  &  3  & $-$7.48& 723\\
Sr {\sc ii} & $-$9.27  & 0.36  &  2  & $-$9.17& 1455\\
Y {\sc ii}  & $-$10.34 & 0.19  &  9  & $-$9.38& 1647 \\
Zr {\sc ii} & $-$9.73  & 0.24  & 16  & $-$9.46& 1736 \\
Ba {\sc ii} & $-$10.68 & 0.09  &  2  & $-$9.86& 1447 \\
Ce {\sc ii} & $\le-$11.22&     &  1  & $-$10.46&1477 \\
Eu {\sc ii} & $\le-$12.37&     &  1  & $-$11.52&1347 \\
\hline
\end{tabular}
\end{table}
 
\begin{figure}
\resizebox{\hsize}{!}{\includegraphics[angle=-90]{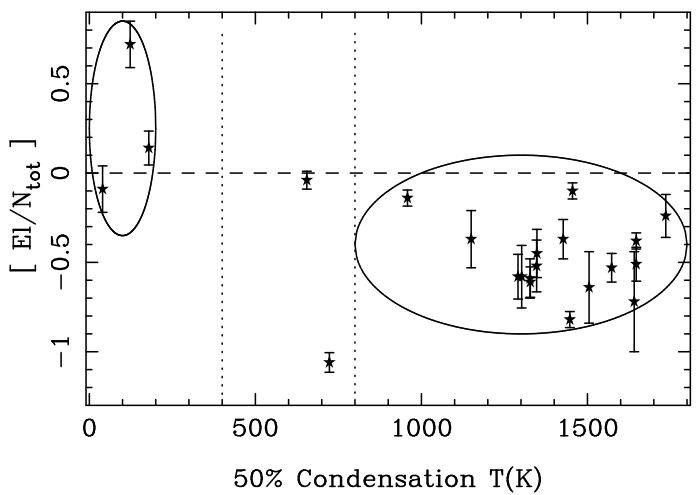}}
\caption{Logarithmic abundance differences vs. 50\%$T_C$.
Error bars are from Table~\ref{tab:abtab}.
Volatile elements, C, N, O are enclosed within the ellipse
on the left; refractory elements fall within the ellipse on
the right.  The intermediate volatiles, Zn and S, are between 
the dashed vertical lines.  The decided outlier, Zn, is discussed
in the text.}
\label{fig:plcond}
\end{figure}

Logarithmic abundance differences--star minus sun--are plotted
in Fig.~\ref{fig:plcond}.
The abundances may reflect a mild $\lambda$ Boo, or 
Vega-like abundance mechanism, where the refractory elements
are depleted while the most volatile elements are 
more nearly normal (Takeda 2008, see also Adelman, et al.
2010, in preparation).  This is not unexpected.
Gray and Corbally (1998) classified the
Herbig Ae star HD 37411 as a $\lambda$ Boo star, while Acke
and Waelkens (2004) describe the Herbig Ae star HD 100546
as a ``clear $\lambda$ Bootis star,'' based on their abundance
analysis.  

Neither
Vega nor HD 101412 would qualify as a $\lambda$ Boo star,
as they do not
meet the 1 dex iron-peak deficiency of Paunzen (2004)
for the $\lambda$ Boo stars.  
Our average deficiency
for the elements with 50\%$T_C > 900$K is about 0.5 dex, half
the depletion for classical $\lambda$ Boo stars.  Nor is
there an obvious trend within these refractory elements 
with 50\%$T_C$.  

Our suggestion that the abundance pattern resembles that of the
$\lambda$ Boo stars rests on average abundances for the 
elements with volatility extremes.  The centroids of the
ellipses in Fig.~\ref{fig:plcond} are significantly displaced
from one another, indicating the most volatile elements are
not depleted while the most refractory elements are.  

\subsection{Outliers\label{sec:outliers}}
The abundance pattern of Fig.~\ref{fig:plcond}  
cannot be explained in terms of
condensation temperature alone.
\subsubsection{Nitrogen\label{sec:nitrogen}}
The high nitrogen abundance may be partially due to 
NLTE.  Kamp, et al. (2001) find nitrogen quite variable among
$\lambda$ Boo stars, and subject to NLTE.  For example, the
LTE abundance excess of N for HD 75654 was +0.65 dex.  The
NLTE calculation reduced this value to +0.30 dex.
\subsubsection{Sodium}
Sodium (50\%$T_C = 953$K), might be considered an intermediate
volatile.  We find its depletion in HD 101412 to be only 0.14 dex.
However, Paunzen, et al. (2002) reported sodium variations from 
$-$1.3 to +1.2 dex with respect to the sun.  Kamp et al. (2008)
discuss this further, and note mechanisms beyond equilibrium
condensation that might account for the scatter.
\subsubsection{Zinc}
Paunzen(2004) listed zinc among the elements {\it expected} to be
deficient in the $\lambda$ Boo stars.  To the extent that we 
suggest the abundances in HD 101412 resemble those of the 
$\lambda$ Boo stars, our finding of an depletion of 1.06
dex ceases to indicate an outlier, and supports the assertion.
 At the same time, it complicates an interpretation
based on volatility.
Surely factors other than equilibrium condensation affect
the stellar abundances of zinc as well as sodium.

\section{Conclusions}

The atmospheric structure of HD 101412 deviates
from a classical model (e.g. ATLAS9).   But the deeper
photosphere, as probed by weaker lines, 
has a line spectrum well produced by
the theoretical model with \teff =  8300K, $\log(g)=3.8$.
Elements with lines from the first and
second spectra yield abundances in good or
satisfactory agreement.  In the worst case, silicon,
Si {\sc i} and Si {\sc ii} disagree by 0.38 dex, only
a factor of 2.4.

The atomic lines are also in general agreement with a mean
{\it excitation} temperature in the range 7200 to 7750K.
An excitation temperature
in this range is expected from
a model with the adopted parameters.

Strong lines show an anomalous saturation, which we have
attributed to maximum line depths, $r_0$, less than that
predicted by the ATLAS9 model.  Similar information comes
from curves of growth, where even the observed depths of the
Mg {\sc ii} $\lambda$4481 doublet (0.53) would lead to a
imaginary value for the microturbulence.

Atomic emission features show departures
of the upper atmosphere from classical.
In addition to emissions seen in the low Balmer
members, broad, generally weak
emissions appear, for example, in [O {\sc i}] $\lambda$6300,
Na {\sc i} D$_1$ and D$_2$, and the O {\sc i}, triplet
$\lambda\lambda$7772, 7774, and 7775.
The widths of these features
are readily measured.  The wavelength spread from the
violet wing to the red wing, of atomic emissions
correspond to velocities of 100 to 200 km s$^{-1}$.
We suggest the upper atmosphere is heated by infalling
material from a primordial disk, and that this heating
is probably responsible for the anomalous saturation.

Most abundances are less than 1 dex from solar values.
With the notable exception of zinc, there is a suggestion
that refractory elements are depleted.  The volatiles
are normal, or in the case of nitrogen, enhanced.

\begin{acknowledgements}
We are especially grateful to Z. Mikul\'{a}\v{s}ek
for communicating the results of his photometric
period in advance of publication and for
an exchange of ideas concerning its interpretation.
It is a pleasure to thank 
J. R. Fuhr J. Reader, and W. Wiese of NIST for advice on
atomic data and processes.
This research has made use of the SIMBAD database, operated
at CDS, Strasbourg, France.
Our calculations have made extensive use of
the VALD atomic data base (Kupka, et al. 1999).
CRC is grateful for advice and helpful conversations
with many of his Michigan colleagues, and to
Jes\'{u}s Hern\'{a}ndez for useful comments and
suggestions.  S. J. Adelman and A. F. Gulliver graciously
shared some results of their forthcoming Vega abundance study.
\end{acknowledgements}

\end{document}